\documentclass[a4paper,fleqn]{cas-dc}
\usepackage[inkscapelatex=false]{svg}
\usepackage{hyperref}
\usepackage[figuresright]{rotating}
\usepackage{color}
\usepackage{algorithm}
\usepackage{algorithmic}
\usepackage[authoryear,longnamesfirst]{natbib}
\def\tsc#1{\csdef{#1}{\textsc{\lowercase{#1}}\xspace}}
\tsc{WGM}
\tsc{QE}
\tsc{EP}
\tsc{PMS}
\tsc{BEC}
\tsc{DE}
\begin{document}
\begin{sloppypar}
\let\printorcid\relax
\let\WriteBookmarks\relax
%\def\floatpagepagefraction{1}
%\def\textpagefraction{.001}
\shortauthors{Gaoqiang Dong et~al.}
\title [mode = title]{An Improved Scheduling with Advantage Actor-Critic for Storm Workloads}
\tnotetext[1]{This work is supported by National Key R\&D Program of China (No.2022ZD0115803), the Scientific Research Foundation of Higher Education (No.XJEDU2022P011), the "Heaven Lake Doctor" project (No.202104120018) and the Doctoral Scientific Research Foundation of Xinjiang University (No.620320029).}
\author[1]{Gaoqiang Dong}[
                        type=editor,
                        auid=000,
                        style=chinese,
                        bioid=1]
% \cormark[1]
\fnmark[1]
\ead{107552101307@stu.xju.edu.cn}
% \credit{Conceptualization of this study, Methodology, Software}
\affiliation[1]{
    %organization={School of Information Science and Engineering},
    addressline={Xinjiang University}, 
    city={Urumqi},
    % citysep={}, % Uncomment if no comma needed between city and postcode
    postcode={830046}, 
    country={China}}
\author[1]{Jia Wang}[
                    type=editor,
                    auid=000,
                    style=chinese,
                    bioid=1]
\cormark[1]
\cortext[cor1]{Corresponding author}
%\cortext[cor2]{Principal corresponding author}
\fnmark[2]
\ead{jw1024@xju.edu.cn}
% \credit{Conceptualization of this study, Methodology, Software}
\author[2]{Mingjing Wang}[
                    type=editor,
                    auid=000,
                    style=chinese,
                    bioid=1]
\cormark[1]
\fnmark[3]
\ead{wangmingjing.style@gmail.com}
% \credit{Conceptualization of this study, Methodology, Software}
\affiliation[2]{
    % organization={},
    addressline={Wenzhou University of Technology}, 
    city={Wenzhou},
    postcode={325000}, 
    country={China}}
\author[1]{Tingting Su}[style=chinese]
% \cormark[1]
\fnmark[4]
% Email ID of the first author
\ead{107552103750@stu.xju.edu.cn}
% \credit{Conceptualization of this study, Methodology, Software}
\begin{abstract}
Various resources as the essential elements of data centers, and the completion time is vital to users.
In terms of the persistence, the periodicity and the spatial-temporal dependence of stream workload, a new Storm scheduler with Advantage Actor-Critic is proposed to improve resource utilization for minimizing the completion time.
A new weighted embedding with a Graph Neural Network is designed to depend on the features of a job comprehensively, which includes the dependence, the types and the positions of tasks in a job.
An improved Advantage Actor-Critic integrating task chosen and executor assignment is proposed to schedule tasks to executors in order to better resource utilization.
Then the status of tasks and executors are updated for the next scheduling.
Compared to existing methods, experimental results show that the proposed Storm scheduler improves resource utilization.
The completion time is reduced by almost 17\% on the TPC-H data set and reduced by almost 25\% on the Alibaba data set.
\end{abstract}

\maketitle 
\section{INTRODUCTION}
\begin{figure}
    \centering
    \includegraphics[width=1.0\columnwidth]{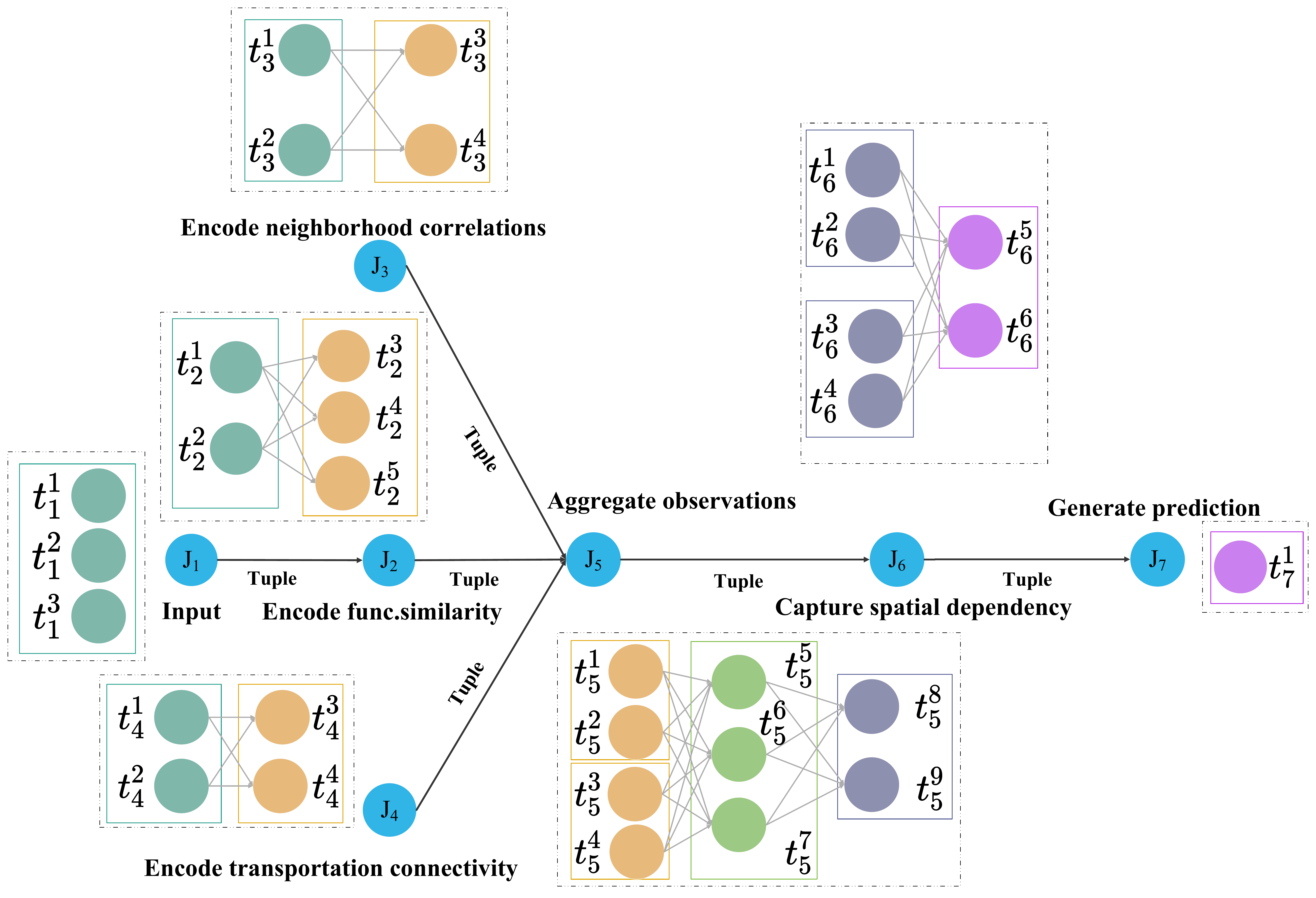}
    \caption{An example of Storm workloads instance.}
    \label{FIG:1}
\end{figure}

With the construction of a smart city, stream data generated by social media, video surveillance and so on, exceeds 30,0000GB per second \cite{eskandari2021scheduler}.
In other words, the Apache Storm\cite{peng2015r} scheduling of stream workloads is paid more and more attention.
Fig.\ref{FIG:1} shows an example of Storm workloads instance.
Generally speaking, traffic stream data are persistent, periodic and spatial-temporal dependent.
The prediction of traffic contains data processing, feature extraction and regression prediction jobs, which are dependent on them.
Some dependent tasks are constructed of a job.
The traffic management department always has its own cluster with some heterogeneous servers.
During the processing of Storm applications, resource managers always focus on resource utilization.
Therefore, it is desirable to design effective and efficient scheduling for the Storm application in order to better resource utilization by minimizing the application completion time.
In this paper, we consider the problem of scheduling Storm workloads to heterogeneous servers in a cluster to minimize the completion time.
Each server has different processing speeds and various numbers of executors.
An executor is the minimal resource unit in Storm.
A lot of Storm applications are processed, each of which consists of many dependent jobs according to the spatial-temporal dependent data.
Meanwhile, the persistent stream data make an application would process until data is finished.
Each application contains multiple jobs, and each job contains some dependent or independent tasks.
The dependent tasks would be processed until its direct processors are finished, while the independent tasks are processed in parallel.

Because of the heterogenous of resources and the complex dependent relationships of jobs with these periodic data, there are two challenges for the problem under study.
(i)The periodic data are generated persistently, i.e., a great discrepancy exists between data sizes of peaks and troughs, which results in a unbalance resource utilization. 
With the growth of applications and big processing data size of peaks, resources would be competed among jobs and resource utilization is high.
While a low resource utilization is obtained with less data size of troughs.
Though existing methods such as fair scheduling \cite{doulamis2007fair}, EMVO \cite{shukri2021enhanced}, and Decima \cite{mao2019learning} are designed for resource utilization, they ignores the unbalance resource utilization between peaks and troughs. 
How to balance the resource utilization among these peaks and troughs is difficult to solve.
(ii)Many jobs would compete for limited resources, which results in the difficult allocation of tasks and executors.
Traffic data with spatial-temporal dependent make dependent jobs be processed subsequently, while independent jobs are processed with resource competing in parallel.
As shown in Fig.\ref{FIG:1}, the job $J_{5}$ would start until jobs $J_{2}$ , $J_{3}$  and $J_{4}$ are finished.
$J_{2}$, $J_{3}$ and $J_{4}$ are competed resource in order to finish almost the same time.
Specially, the heterogenous servers leads more difficult to assignments of tasks and executors in peaks.
The persistent, periodic and spatial-temporal dependent stream data make traffic applications contain many dependent jobs and make stream workloads fluctuate among peaks and troughs.

With the limited resource in a cluster, the efficient Storm scheduling is needed to be proposed.
In this paper, we picture a weighted embedding with Graph Neural Network (GNN) to capture global features of all dependent jobs according to the dependencies, the types and the processing servers of jobs.
With the advantage of stability, A2C is adopted to assign chosen tasks to reasonable executors in order to avoid local optimal solution for periodic traffic stream.

The main contributions of this paper are summarized as follows:
\begin{itemize}
\item A new weighted embedding with GNN is presented to extract features of jobs comprehensively in order to improve assignments among tasks and executors for resource utilization, especially in peaks and troughs.
\item An enhanced Advantage Actor-Critic consisting of task selection and executor allocation is proposed to better assignments. 
With the obtained assignments of the actor network, the critic network is used to measure these assignments by an evaluation function.
Meanwhile, the parameters of the actor network are updated.
\end{itemize}

The rest of this paper is organized as follows: Section 2 reviews related works.
A description of the considered problem is presented in Section 3. 
Section 4 describes a Storm scheduling strategy for the studied problem.
Experimental results are shown in Section 5 followed by conclusions and future work in Section 6.

\section{RELATED WORK}
Recently, some studies have been focused on Storm workload task scheduling, which allocates tasks to executors.
Because of the continuously generated periodic data, imbalanced resource utilization emerged because of the data sizes between peaks and valleys.
Furthermore, streaming data have spatiotemporal dependencies, which means that many tasks will compete for limited resources.
This will lead to an unfair allocation of resources.
Therefore, an effective scheduling method is needed to balance resource utilization and reduce completion time.
In terms of balanced resource utilization, there are two kinds of scheduling problems with resource management and fair allocation of resources, respectively.
And, optimizing completion time impacts resource utilization.
So, the detailed works on resource management and task scheduling for unfair resource allocation are described in this section.

Much attention has been paid to resource management. 
Tasking scheduling and resource provisioning can efficient use of resource.
Many algorithms have been used to solve resource management problems.
These algorithms all try to reducing completion times, making more efficient use of resource, and achieving load balancing.
Among them, \cite{muhammad2021ban}et al propose BAN-Storm.
BAN-Storm is a stream scheduler based on  resource-aware mapping mechanism, that calculates the computing power of each executor and then assigns task groups to the execut.
Aimed at achieving higher resource utilization in Storm systems.
However, because of Storm processes periodic data, the application will experience different load levels at different times.
The resource utilization is low except the peak hours.
\cite{pang2023pac} et al proposed a heterogeneous preference-aware collaborative scheduling scheme(PAC).
PAC ensures the required QoS of latency-critical applications, and also improves throughput.
Since the real-time arrival of multi-source high-speed data streams, it brings challenges to resource management.
\cite{sun2021lr} et al introduced Lr-Stream, a scheduling framework that is both delay- and resource-aware. 
This framework led to substantial enhancements in system latency reduction and increased overall system throughput.
However, when dealing with intricate traffic scenarios and balancing resource utilization, reinforcement learning appears to be more advantageous.
\cite{mao2019learning} et al first proposed to use reinforcement learning to solve cluster resource scheduling problems, which laid the foundation for subsequent research.
Subsequently, \cite{asghari2021task} et al the combination of SARSA and genetic algorithm is used to manage resources.
In other words,  this is conducted by selecting the most appropriate set of the tasks that maximizes the utilization of the resource.
\cite{islam2021performance} et al. proposed the TF-Agents framework, which contains two schedulers based on deep reinforcement learning (DRL).
The DRL-based scheduling agent matches tasks to job executors at a fine-grained level, improving resource utilization.
These approaches are designed to evaluate resource needs and manage resources, they also contribute significantly to enhancing resource utilization.
Nevertheless, they overlooks concerns unbalanced resource utilization and extended completion times resulting from disparities in data size during the processing of real-time periodic flow data peaks and troughs.

In addition, different applications require access to different heterogeneous resources. 
At the same time, the inherent spatiotemporal dependence of traffic flow data increases the difficulty of selecting tasks and allocating resources.
It results in task competition for resources.
This brings about problems such as unfair resource allocation, inaccurate resource requests, and unsatisfactory resource utilization.
Running deep learning jobs shared on GPUs compete for GPU resources, thereby reducing the overall performance and efficiency of the system. 
To enhance resource utilization and address issues of unfair resource allocation. \cite{yeung2021horus} et al propose an interference-aware and prediction-based resource manager for deep learning systems(Horus).
Horus proactively predicts GPU utilization of heterogeneous deep learning jobs, removing the need for online profiling and isolated reserved GPUs.
However, more researchers are committed to exploring various task scheduling strategies to address the issues of fairness and efficiency in resource allocation.
\cite{jin2020improving} et al designed a new framework, called Ursa, which enables the scheduler to capture accurate resource demands dynamically and provide timely, fine-grained resource allocation based on monotasks.
This method can effectively improve resource utilization.
But, because of the online scheduling process of dependent tasks, optimal resource utilization cannot be guaranteed.
\cite{asghari2020online} et al employed reinforcement learning techniques for both task scheduling and resource allocation. 
In the process of task scheduling and resource allocation, besides resource contention, the execution order of tasks also impacts the job completion time.
\cite{duan2020reducing} et al propose a reinforcement learning-based scheduler.
It can adaptively adjust the resource contention.
In various application scenarios \cite{talaat2022effective}, \cite{8959320}, \cite{meyer2021ml} and \cite{hu2019spear}.
The focus is on selecting the most suitable task sets and efficiently matching them with available resources to maximize resource utilization. 
Extending prior research, \cite{xia2022multi} et al propose a genetic algorithm (GALCS) based on the longest common subsequence (LCS) to solve the allocation and scheduling problems of different resources.
While the mentioned approaches take into account the influence of dependent tasks on resource utilization.
They overlook the intrinsic attributes of streaming data and the intricacies associated with task selection and resource allocation within constrained resource environments.

In other words, these researches fail to tackle the crucial choices regarding the prioritization and assignment of tasks to particular actuators.
Compared to the existing scheduling with the problem considered in this paper is different in two aspects:
(i)Since users care more about completion time and resource utilization is affected by completion time ,the objective of this paper is to minimize completion time and enhance the overall performance of storm clusters.
(ii)Because of the unfair resource allocation increased completion times of the entire application, an improvement of assignments of tasks and executors, an improvement of  is also considered in this paper.
In this paper, we present a deep reinforcement learning model designed to address issues related to resource utilization imbalance and competitive resource allocation.

\section{PROBLEM DESCRIPTION}
In this paper, the directed acyclic graph (DAG) is allocated to a cluster of M servers. A DAG is a Storm application which consists of many dependent jobs $J = \left \{ J_{1}, J_{2}, \cdots, J_{i}, \cdots , J_{n} \right \}$. 
Each job is comprosed of n tasks,  $J_{i} = \left \{ t_{i}^{1}, t_{i}^{2},\cdots, t_{i}^{n} \right \}$. 
The execution time and data volume of each task are represented by $P_{n}$ and $D_{n}$. 
In a dynamic real-time streaming environment, the future task execution times $P_{n}$ for different jobs are unknown.
The servers is represented as $R = \left \{ R_{1}, R_{2},\cdots, R_{k},\cdots, R_{M} \right \}$, where each server $R_{k}$ contains $R_{k} = \left \{ e_{k}^{1}, e_{k}^{2},\cdots, e_{k}^{m} \right \}$ executors. 
Because of the application is continuously processed, and the number of data processed has peaks and troughs.
Given the limited number of executors in the cluster.
Our objective is to propose a scheduling policy that optimally balances resource utilization, and allocates resources to jobs in a manner that minimizes the completion time.

Due to the periodicity of streaming data, a peak and trough of processing data occurs. 
At certain times, the number of tasks may be larger and resource requirements can be higher, while at other times, the workload may be smaller.
This can cause tasks to take too long to complete, and imbalance in resource utilization occurs.
Every task demands optimal resource utilization.
Unfair resource allocation increased completion times of the entire application.
Therefore, in the paper, we balance resource utilization by reducing completion time. 
The specific formula is as follows:
Due to the periodicity of streaming data,  a peak and trough of processing data occurs.
At certain times, the number of tasks may be larger and resource requirements can be higher, while at other times, the workload may be smaller.
This can cause tasks to take too long to complete, and imbalance in resource utilization occurs.
Every task demands optimal resource utilization.
Unfair resource allocation reduced completion times of the entire application.
Therefore, in the paper, we balance resource utilization by reducing completion time. 
The specific formula is as follows:
\begin{equation}
U=\frac{T}{e_{k}^{n} \times C_{ik}^{t}} 
\end{equation}

where $U$ represents resource utilization and $C_{ik}^t$ represents the completion time of task $t_{i}^{n}$ assigned to executor $e_{k}^{n}$ for job i, the overall completion time is influenced by the individual task completion times.
T denotes the total executors time used by the tasks. $e_{k}^{n}$ represents the number of allocated executors. 
A smaller $C_{ik}^t$ accelerates job processing, thereby enhancing resource utilization.

In a storm cluster, our goal is to enhance streaming job performance by minimizing the completion time:
\begin{equation}
\min \sum_{i=1}^{N} \sum_{n=1}^{t_{i}^{n}} \sum_{k=1}^{e_{k}^{M}} x_{ik}^{t}C_{ik}^{t}
\end{equation}

where \(x_{ik}^t\) denotes whether task $t_{i}^{n}$ of job i is allocated to executor k on a server (with \(x_{ik}^t\) being a binary variable, equal to 1 for successful allocation and 0 otherwise).

To ensure a balanced resource utilization and minimize the system's average workload, we objective to optimize the completion time, guided by Little's law [22]:
\begin{equation}
\begin{split}
    C_{ik}^{t} = \min W_{ik}^{t},\forall i\in \left \{ 1,2,3,\cdots,N \right \}, \\ 
    n\in \left \{ 1,2,3,\cdots, t_{i}^{n}\right \},\\
    k\in \left \{ 1,2,3,\cdots, e_{k}^{m}\right \}
\end{split}
\end{equation}

where,\(W_{ik}^t\) denotes the waiting duration for the $t_{i}^{n}$ tasks of job i when it is allocated to executor $e_{k}^{m}$.

\textbf{Start time limit}: Stream jobs can initiate transmission only after the executor releases them. So:
\begin{equation}
     \overline{F_{n}} \ge e_{n}  \forall n\in \left \{ 1,2,\cdots ,N \right \}
\end{equation}

Where, $\overline{F_{n}}$ represents the initial transmission time of task $t_{n}$. 
When $\gamma _{n} \le  e_{n}$, $e_{n}$ represents the release time of the executor after execution.

\textbf{Dependency constraints}: Jobs (referred to as $J_{n}$) possess dependencies. The subsequent stages should be executed once the preceding stage is completed.
Thus, it's essential to address jobs with interdependencies.
So, the dependency constraint is expressed as:
\begin{equation}
\begin{split}
    \overline{F_{n}}\ge S_{n}j_{n} \forall n\in \left \{ 1,2,\cdots ,N \right \}
\end{split}
\end{equation}

Where $S_{n}$ represents a binary indicator when $j_{n}$ is dependent, it is represented as 1, otherwise, it is represented as 0.

\section{PROPOSED ALGORITHM}
\label{4}
With cyclical data continuing to be generated, the uneven data size between peaks and troughs leads to unbalanced resource utilization.
And because of the dependencies between tasks, tasks will compete for resources, causing the problem of unfair resource allocation.
In terms of \cite{buddhika2017online}, it is not hard to derive that the considered optimal scheduling of stream processing computations is the resource-constrained scheduling problem, and can be characterized as either NP-Hard.
DRL is effective for NP-hard combinational optimization problems.
In this paper, we propose an efficient  adaptive task scheduling framework(ATSF) based on DRL. 

The ATSF framework is shown in Algorithm \ref{alg1}.
The comprises five key components: the GNN-weighted embedding module, the Enhanced Advantage Actor-Critic network module, the DPL module, the SASJ module, and the UAEAJ module. 
Our objective is to minimize the completion time by addressing the issues of resource utilization imbalance and competitive resource allocation.
To achieve this, we utilize Apache Storm's real-time, low-latency data stream processing capabilities and harness the characteristics of streaming data to propose an adaptive task scheduling method based on DRL.
Initially, we begin by processing traffic flow data within the road network using GNN-weighted embedding module. Subsequently, in DPL\ref{alg2}, we utilize priority scores generated by the Enhanced Advantage Actor-Critic network. 
These scores encompass both task selection and execution matching priorities, forming a priority policy list. 
When a task is completed or an application is released, SASJ \ref{alg3} is invoked to schedule tasks based on a priority policy list. 
UAEAJ \ref{alg4} then updates the pool of available executors and the list of scheduling active tasks.
It's worth noting that all executors start in an idle state initially. 
Finally, the scheduling framework we have designed is trained using TSA\ref{alg5}.
Subsequently, we will introduce the remaining components one by one.

\begin{algorithm} 
	\caption{Adaptive Task Scheduling Framework (ATSF)} 
	\label{alg1} 
	\begin{algorithmic}[1]
		\REQUIRE A new DAG contain with jobs: $J_{n}$ 
		\ENSURE The scheduling priority for the active task set $t_{i}^{n}$ and executors $e_{k}^{m}$
		\STATE Initialize a list of all available executors: $e_{k}^{m}$ in each server $R_{M}$
        \FOR{i =1 to n}
        \STATE Deal with the job $J_{i}$ of DAG using the GNN Weight Embedding $H_{v}^{i}$
        \STATE Calculate $H_{v}^{i}$,$y^{i}$,$z$ using Equation(6)
        \ENDFOR
        \REPEAT
        \STATE Calculate the priority score ($q_{t}^{i}$) and executor parallelism score ($w_{l}^{i}$) using Enhanced Advantage Actor-Critic
		\STATE Determining scheduling priority list P using DPL
        \STATE Call SASJ to allocate tasks in P to executors in $e_{k}^{m}$
        \STATE Updating the available executors $e_{k}^{m}$ and Set of schedulable tasks $t_{i}^{n}$ in an $J_{i}$ for UAEAJ
        \UNTIL{All tasks $t_{i}^{n}$in $J_{n}$ being Scheduled}
        \RETURN
	\end{algorithmic} 
\end{algorithm}

\subsection{GNN weighted embedding}
A DAG is composed of many dependent jobs $J = \left \{ J_{1}, J_{2}, \cdots, J_{i}, \cdots, J_{n} \right \}$. 
Each job $J_{i}$ consists of n tasks represented as, $J_{i} = \left \{ t_{i}^{1}, t_{i}^{2},\cdots, t_{i}^{n} \right \}$. 
Traditional graph traversal algorithms, such as depth-first search (DFS) or breadth-first search (BFS) traverse the DAG and obtain information about topology.
However, these methods cannot extract the global features of DAG, ignore the correlation between long-distance jobs, and have limitations in adapting to different graph structures.
Instead, we utilize graph neural networks to process incoming DAGs.
Unlike traditional methods,  our focus is on capturing details like task priority, resource usage, and job arrival rate. 
We input an new DAG to generate embedding vectors through the embedding process. 
These vectors effectively represent task count, task execution time, and parallelism limit within the node. 
The embedding process produces three separate embedding vectors: job embedding, DAG embedding, and global embedding. 
This approach provides a thorough representation of the state messages while also prioritizing scalability and efficiency.

\textbf{Job embedding}: Record details about the present node and its subordinate nodes (such as those along the critical path involving the target node), and express this information in a mathematical format.
\begin{figure}
  \centering 
  \includegraphics[scale=0.15]{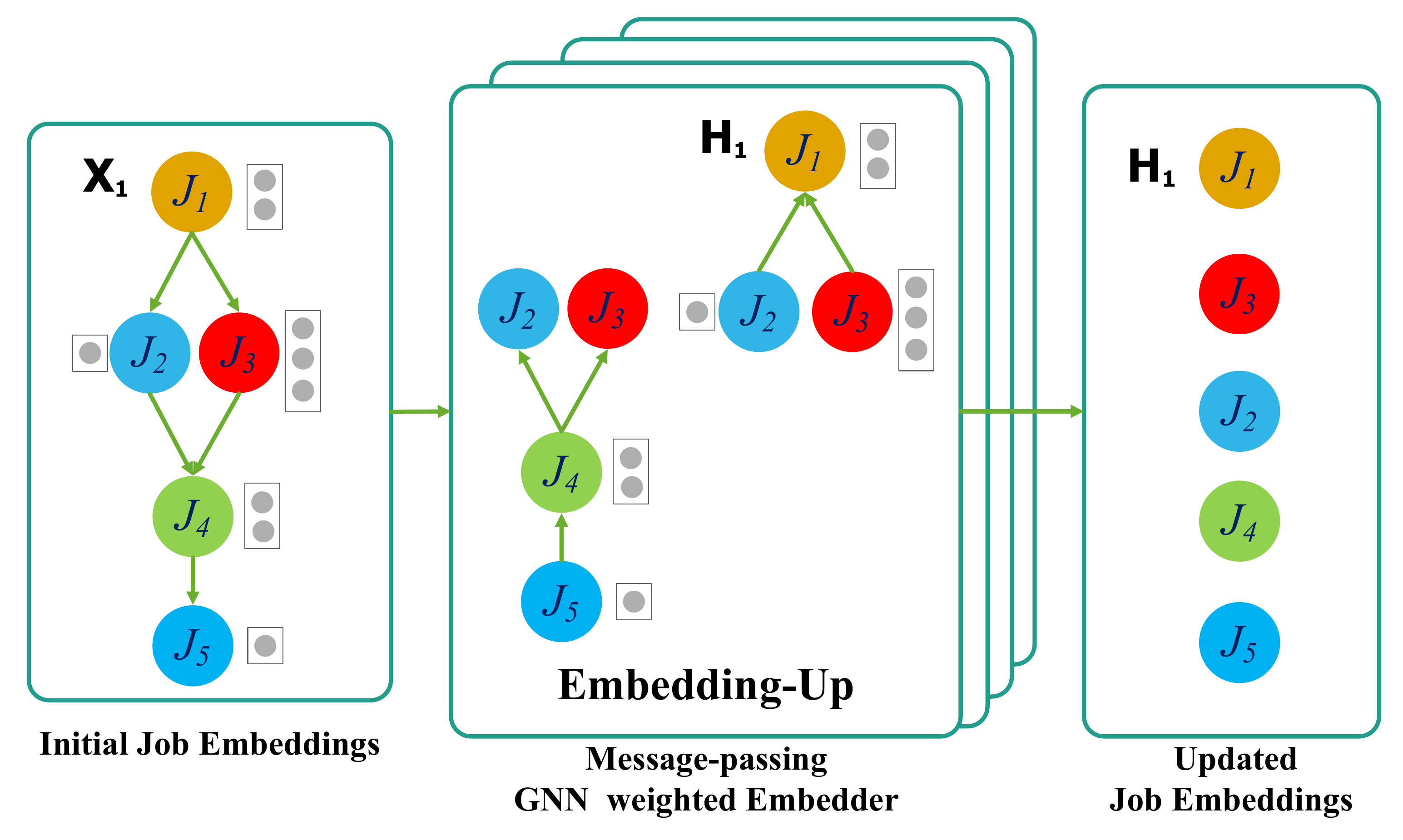}
  % \captionsetup{justification=centering} % 设置图标题居中
  \caption{Job embedding}
  \label{FIG:2}
\end{figure} 
We create an embedding vector, denoted as $ H_{v}^{i} $, for each node in the DAG $ J_{i} $ using corresponding jobs attribute vectors $ x_{j}^{i} $. 
This embedding vector encapsulates information about all jobs that can be reached from job j, including its child jobs and other descendants within the DAG. 
The scheduler derives these vectors by traversing the DAG starting from the child jobs of j, as depicted in Fig.\ref{FIG:2}.
This process involves sequentially visiting nodes within the job until all jobs reachable from j have been covered. Subsequently, through several message-passing iterations, information is gradually communicated from leaf nodes to parent nodes.
This propagation culminates in the computation of embedding vectors for an job, following the provided formula.
\begin{equation}
H_{j}^{i} = g\left[\sum_{u=\xi \left ( j \right )}\alpha_{uj}^{i}\odot f\left ( H_{u}^{i} \right ) \right ]+x_{j}^{i}
\end{equation}

Then, We draw inspiration from \cite{mao2019learning} to introduce two identical nonlinear transformations, denoted as $f(\bullet)$ and $g(\bullet)$. These transformations take the form of a compact neural network and are applied uniformly to all jobs embedding and messages during the transfer step.
Furthermore, we incorporate a weight coefficient that considers the task volume in the embedding process, denoted as $\alpha_{uj}^{i} \left ( H_{j}^{i}, H_{u}^{i-1} \right) = Softmax_{u=\xi \left ( j \right )}\left ( H_{j}^{i}, H_{u}^{i-1} \right )$.
Unlike conventional graph neural networks that utilize the aggregation operation $H_{j}^{i} = \sum_{u=\xi \left ( j \right )}f\left ( H_{u}^{i} \right)$, the incorporation of dual nonlinear transformations enhances the ability of graph neural network to compute features such as the critical path of a DAG.
This approach enables efficient maximum operations among jobs in a clustered environment. The calculation method for DAG embedding $y^{i}$ and global embedding $z$ remains the same as described above.

Finally, we use embedding vectors as input to the Enhanced Advantage Actor-Critic network module.
And introduce two scoring metrics into the A2C network: the priority score of a task denoted as $q_{t}^{i}$ and the parallelism score denoted as $w_{l}^{i}$.
Both indexed by task/parallelism identifiers. 
By multiplying these scores, we create a composite priority list denoted as $P = \left \{ P_{1}, P_{2}, P_{3},\cdots, P_{\left | t \right | \times  \left | l \right | } \right \} $, representing the scheduling strategy. 
In this list, $P_{\left | t \right |}$ represents the priority value of a schedulable task for application $J$, and $P_{\left | l \right |}$ represents the priority value of the parallelism level of executor.
Higher values of $P_{t}$ and $P_{l}$ correspond to higher priorities.
Please refer to DPL.\ref{alg2} for a detailed explanation.

\begin{algorithm} 
	\caption{Determining Priority List (DPL)} 
	\label{alg2} 
	\begin{algorithmic}[1]
		\REQUIRE job embedding  $H_{j}^{i}$, DAG embedding  $y^{i}$ and global embedding $z$, the number of executors $e_{k}^{m}$, the number of tasks $t_{i}^{n}$,Parallelism $l$,task $t$
		\ENSURE Priority list P of scheduling active tasks and executor
        \STATE Training based on Advantage Actor-Critic for TSA
        \FOR{t =1 to $t_{i}^{n}$ }
        \STATE Calculate $q_{t}^{i}$ using $q_{t}^{i} \triangleq q\left (H_{t}^{i},y^{i},z\right)$
        \ENDFOR
        \FOR{l = 1 to $e_{k}^{m}$}
        \STATE Calculate $w_{l}^{i}$ using $w_{l}^{i} \triangleq w\left(y^{i}, z, l\right)$ 
        \ENDFOR
        \STATE Calculate the probability of selecting task $P_{\left | t \right |}$ using Equation (7)
        \STATE Calculate the probability of Executor parallelism selection $P_{\left | l \right |}$
        \RETURN priority list $P = \left \{ P_{1},P_{2},P_{3},\cdots ,P_{\left | t \right | \times  \left | l \right |   } \right \}$
	\end{algorithmic} 
\end{algorithm}

In DPL \ref{alg2}, Line 3 defines the priority score ($q_{t}^{i}$) of the computing task and Line 4 defines the executor parallelism score ($w_{l}^{i}$).
During application execution, if the number of executors is below the parallelism limit ($l_{i}$), the scheduler will assign extra executors to DAG until the limit ($l_{i}$) is attained. 

\textbf{Tasks selection}: In the scheduling event at the current time T, the policy network selects one of the applications to start scheduling. For a task $t_{i}^{n}$ in the $J_{i}$, the Policy Network calculates the score $q_{t}^{i} \triangleq q\left ( H_{t}^{i},y^{i},z\right) $, where $q\left( \cdot \right ) $ is the scoring function that maps the embedding vector to a scalar value. The scalar value is used to measure the priority of each task, and the scoring function is implemented by the nonlinear transformation of the neural network. Then the softmax operation \cite{nasrabadi2007pattern} calculates the probability of selecting task $t_{i}^{n}$ based on the priority score:

\begin{equation}
P(\text {task}=t)=\frac{\exp \left(q_{t}^{i} \right)}{\sum_{u \in \mathcal{A}_T} \exp \left(q_u ^{j(u)}\right)}
\end{equation}

Where j(u) represents the jobs u of task $t_{u}^{n}$, and $A_{t}$ represents the set of tasks that can be scheduled at time T. This set stipulates that Softmax only needs to calculate the probability of scheduling active tasks.

The existing scheduler assigns a fixed degree of executor to each task. 
For instance, in the Storm cluster environment, the "parallelism hint" parameter (indicating the initial number of an executor) at application submission determines the overall degree of parallelism for the application. 
However, we are now adopting a dynamic approach to adjust the parallelism of the application degree based on the requirements of its different stages. 
As various tasks within the application become runnable or complete their execution, the degree of parallelism adjusts accordingly.
The probability of choosing a specific parallelism limit $P\left(l \right )$ is calculated in the same way as $P\left(t \right )$.

\subsection{Scheduling Active Streaming Jobs}
The objective of SASJ.\ref{alg3} is task-to-executor mapping based on the priority list P obtained from DPL.\ref{alg2}.
In other words, it decides which executor handles which task in the Storm cluster. 
To begin scheduling, we must first verify if there are any idle executors in the cluster. If there aren't any, we should release some executors.
The time complexity can be expressed as $O(N)$, where $N$ is the total number of active tasks.
\begin{algorithm} 
	\caption{Scheduling Active Streaming Jobs (SASJ)} 
	\label{alg3} 
	\begin{algorithmic}[1]
		\REQUIRE Priority list P of active tasks and executors
		\ENSURE Specific scheduling decisions
        \FOR{for $t_{i}^{n}$ in $J_{i}$}
        \STATE Retrieve the idle executors at the moment: $e_{k}^{m}$
        \FOR{each unfinished $t_{i}^{n-1}$ in $J_{i}$} 
        \STATE allocate executor of $e_{k}^{m-1}$ to $t_{i}^{n-1}$
        \STATE Updating the available executors $e_{k}^{m-1}$ for UAEAJ
        \ENDFOR
        \FOR{each finished $t_{i}^{n-1}$ in $J_{i}$}
        \STATE release executors $e_{k}^{m-1}$
        \STATE Updating the available executors $e_{k}^{m-1}$ for UAEAJ
        \ENDFOR
        \ENDFOR
        \STATE allocate remaining executors equally to all jobs
        \RETURN specific scheduling decisions
	\end{algorithmic} 
\end{algorithm}

Once an executor is assigned to a task.
UAEAJ.\ref{alg4} serves to refresh the active of tasks and the pool of accessible executors.
These executors in Storm are tailored to handle akin tasks within the directed acyclic graphs (DAGs). 
The dynamic allocation of executors aligns with scheduling determinations and the replenishment of executors for new tasks upon the culmination of ongoing ones remains pivotal for upholding system performance. Additionally, Tasks are performed within a specific application, the list of tasks scheduled for subsequent execution is continually adjusted to ensure compliance with dependency constraints.

\begin{algorithm} 
	\caption{Updating the Available Executors And Jobs (UAEAJ)} 
	\label{alg4} 
	\begin{algorithmic}[1]
		\REQUIRE List of scheduled executors $e_{k}^{m-1}$ and tasks $t_{i}^{n-1}$
		\ENSURE List of available executors $e_{k}^{m}$ and tasks $t_{i}^{n}$
        \IF{allocate executor of $e_{k}^{m-1}$ to $t_{i}^{n-1}$ }
        \STATE update $e_{k}^{m}$ by deducting $e_{k}^{m-1}$ from them
        \ELSIF{release executor of $e_{k}^{m-1}$}
        \STATE update $e_{k}^{m}$ by adding $e_{k}^{m-1}$ from them
        \ENDIF
        \FOR{each finished task $t_{i}^{n-1}$ in $J_{i}$}
		\STATE update $t_{i}^{n}$ by deducting $t_{i}^{n-1}$ from them 
        \ENDFOR
        \RETURN List of available executors $e_{k}^{m}$ and tasks $t_{i}^{n}$
	\end{algorithmic} 
\end{algorithm}

\subsection{A2C method used to train Scheduling Agent}
We present the pseudocode for training the Scheduling Agent.
The policy network takes the embedding vector obtained from data processing as input and produces a two-dimensional action denoted as $\left \langle t, l_{i}\right \rangle $. 
These action values are limited by the executor parallelism constraints of task t, and $l_{i}$. 
The training process commences by randomly selecting an episode length denoted as $\mu$ from an exponential distribution, as indicated in line 4.
The initial $\mu_{mean}$ is deliberately kept low, but its average length is gradually increased to avoid inefficient training time. 
Subsequently, N experience fragments are gathered by sampling the job space (line 5). 
To mitigate variability arising from random job arrivals, the Generalized Advantage Estimation (GAE) technique is used to calculate the advantage function (line 11). 
Line 13 describes the A2C algorithm, which updates the Scheduling Agent's strategy parameter $\theta$.

\begin{algorithm} 
	\caption{A2C method used to train Scheduling Agent (TSA)} 
	\label{alg5} 
	\begin{algorithmic}[1]
		%\REQUIRE $n \geq 0 \vee x \neq 0$ 
		%\ENSURE $y = x^n$ 
		\STATE Initialize all neural networks
        \FOR{enach iteration}
		\STATE Sample episode length $\mu$ $\sim$ $exponential\left ( \mu _{mean} \right ) $ 
		\STATE Sample a job arrival sequence
		\STATE Run episodes $i =1,\cdots ,N$: 
        \STATE $\left \{  s_{1}^{i}, a_{1}^{i},r_{1}^{i},\cdots , s_{\mu }^{i}, a_{\mu }^{i},r_{\mu}^{i}\right \}\sim \pi_{\theta }$
        \STATE Compute total reward: $R_{k}^{i} =  {\textstyle \sum_{k'=k}^{\mu}r_{k'}^{i}} $
        \FOR{$k = 1$ to $\mu$}
        \FOR{$i = 1$ to N}
        \STATE compute advantage function(GAE):
        \STATE $A\left ( s_{k}^{i},a_{k}^{i}\right ) = R_{k}^{i} + \gamma V^{\pi}\left ( s_{k+1}^{i} \right ) - V^{\pi}\left ( s_{k}^{i} \right )$
        \STATE $V^{\pi} = \lambda ^{i}R_{k+1}^{i}$
        \STATE $\bigtriangleup \theta \leftarrow \bigtriangleup \theta + \nabla_\theta \left (  \log \pi_\theta\left(s_{k}^{i}, a_{k}^{i}\right) \right)A\left ( s_{k}^{i},a_{k}^{i}\right ) $
        \ENDFOR
        \ENDFOR
        \STATE $\mu _{mean} \leftarrow \mu _{mean}+\varepsilon $
        \STATE $\theta \leftarrow \theta  + \alpha \bigtriangleup \theta  $
        \ENDFOR
	\end{algorithmic} 
\end{algorithm}

The TSA.\ref{alg5} operates as follows: denoted as $\pi_{\theta} = (s_{k}, a_{k})$, where $a_{k}$ provides the current state $s_{k}$ with a prioritized list of ongoing scheduling active tasks and a maximum executor limit. 
After each action $a_{k}$, the Scheduling Agent receives a reward $r_{k}$, evaluating the action's quality relative to the optimal goal. 
The reward is computed as $r_k = -\left ( t_{k} - t_{k-1} \right ) n_{k}$, with $t_{k}$ as the time after the $k^{th}$ action, and $n_{k}$ as the tasks executed in job $J_{i}$ during $\left [ t_{k}, t_{k-1} \right )$.
The objective within each episode is to maximize the cumulative reward ${\textstyle \sum_{k=0}^{A}r_{k}}$, where A is the total actions taken in the episode.

In a continuous streaming scenario, the main goal is to effectively handle a model amidst a constant flow of tasks. 
However, the variations in task arrivals directly impact the rewards received by the Agent.
When fewer tasks arrive, the Agent experiences slight penalties, whereas a sudden influx of task arrivals leads to greater penalties due to growing task queues. 
This fluctuation in rewards occurs irrespective of task execution time and is solely due to the random nature of task arrivals. 
This disparity not only brings uncertainty to the optimal approach but also introduces noise that obstructs efficient training.

While the A2C algorithm utilizes a value function to estimate cumulative rewards for states over the long term, it struggles to fully mitigate the unpredictability of rewards within an episode. 
Yet, the unpredictable nature of task arrivals presents an additional challenge.
To address this concern, we embrace a strategy inspired by reference \cite{mao2018variance} that focuses on inputs to minimize this discrepancy. 
This approach involves averaging episodes with different sequences of task arrivals and, during training, randomly selecting working sequences across multiple iterations. 
This method effectively reduces the fluctuations in task arrival patterns, enabling the A2C algorithm to formulate more precise scheduling strategies.

\section{EVALUATION}
We evaluated our model's performance on the Huawei cloud testbed and simulated storm cluster using a workload from the TPC-H data set and  Alibaba cluster trace. 
The number of Storm applications is set as {1,22,44}.
For each applications, the number of DAG-dependent Storm jobs is generated with TPC-H and Alibaba data set. 
In other words, We generate our workload using the TPC-H \cite{chiba2016workload} data set, creating data streams from 22 queries ($Query_{1}\sim Query_{22}$).
These streams include various information like job and task counts, execution times, and executor requirements. 
According to the scale factor, data of 2G, 5G, 10G, 20G, 50G, 80G and 100G are generated respectively.
In addition, we generate our workload from Alibaba cluster trace \cite{Alibaba2018clustertrace}, creating data streams from 22 Jobs(containing multiple tasks and instances). 
After slicing and partitioning the data, we organize it into sizes of 2G, 5G, 10G, 20G, 50G, 80G, and 100G.
Finally, We combined the TPC-H and Alibaba data sets in equal proportions to create a comprehensive streaming DAG collection, comprising 44 distinct applications. 
This integrated data set has been segmented into data partitions varying in size from 2 GB to 100 GB, catering to diverse application requirements.
Both the training and test sets for the experiments are randomly extracted from the data partition, and they share an identical internal distribution.
This mixed data set will be utilized for parameter verification.
This TPC-H and Alibaba data set will be utilized for algorithm comparison.

Then we randomly sample from these data areas to create diverse data streams. 
These streams are used to form different topologies, and from there, we construct a DAG (Directed Acyclic Graph) for streaming jobs by generating an adjacency matrix.
Each job DAG comprises 'n' tasks and randomly generated dependencies (n-1). Job arrivals follow a Poisson distribution with the workloads being $WLD = 40\%$. 
To parallelize episodes with the same job sequence, we employ 16 agents.
In addition, We utilize TensorFlow\cite{pang2020deep} to construct our model's training framework within a simulation environment. And ,we employ Python 3.7 to develop a Storm scheduling simulator. This entire setup operates on the Ubuntu 16.04 operating system. In Huawei Cloud, we rent two machines of size 1*NVIDIA T4/1*16G and 1*NVIDIA V100-SMX2/1*16G for simulation experiments.
Subsequently, we evaluate our model by minimizing the completion time.

\subsection{Parameters Tuning}
In terms of the ATSF framework show in Sect.\ref{4}, five components(GNN-weighted embedding, Enhanced Advantage Actor-Critic network, DPL, SASJ, and UAEAJ have a large expected impact on the performance.
Different workloads lead to different performance.
Let $WLD$ denote the different workloads.
We respectively compared the complete time under 20\%, 40\%, 60\%, and 80\% workloads.
Various learning rates and hidden dimensions lead to different performances too.
$Hd$ is the hidden dimension, which is used in enhanced advantage actor-critic network.
We test $Hd$ with four representative values $\left ( 8,4,2,1 \right )$, $\left ( 16,8,4,1 \right )$, $\left ( 32,16,8,1 \right )$, $\left ( 64,32,16,1 \right )$.
Let $Hd_{8}$, $Hd_{16}$, $Hd_{32}$, and $Hd_{64}$ respectively represent $Hd$ with $Hd_{8} =\left ( 8,4,2,1 \right ) $, $Hd_{16} =\left ( 16,8,4,1 \right ) $, $Hd_{32} =\left ( 32,16,8,1 \right )$, $Hd_{64} =\left ( 64,32,16,1 \right )$.
$\alpha$ and $\beta$ are the learning rates of the Actor network and Critic network.
Let $\ell_{1} $, $\ell_{2} $, $\ell_{3} $, $\ell_{4} $ with four representative values $\ell_{1}  = \left ( \alpha = 0.01, \beta = 0.1 \right )$, $\ell_{2} = \left ( \alpha = 0.001, \beta = 0.01 \right )$, $\ell_{3} = \left ( \alpha = 0.0001, \beta = 0.001 \right )$, $\ell_{4} = \left ( \alpha = 0.005, \beta = 0.05 \right )$.
Since the considered discount factor $\gamma$ affects strategies.
So, we test  $\gamma$ with four representative values $\gamma_{1} = 0.97 $, $\gamma_{2} = 0.98$, $\gamma_{3} = 0.99$, $\gamma_{4} = 1.01$.
Different the number of agents result in different completion times of jobs.
Four the number of agents(4, 8, 16, 32) are compared.
In addition, we compare the maximum depth of root-leaf message passing ($max_{depth}={2,4,8,16}$) in our models.
As a result from all the above options, there are $4 \times 4\times 4\times 4\times 4 = 1024$ treatments in the experimental design of parameters tuning.
With the aforementioned four workloads, the number of treatments is up to 4096. 
Each treatment are tested five random instances, and the total number of results in the calibration experiment is 20480. All test sets are generated by mixed data set.

Experimental results are show in Table\ref{tbl:table1}: we validated five of parameters and examined their impact on completion time.
First, as shown in Table, we compare the impact of hidden dimensions in ATSF on experiments.
Through experimental comparison, it is proved that (32, 16, 8, 1) performs best.
Afterwards, as shown in Table, we compared the impact of different learning rates on the experiments in TSA.
Through experimental comparison, it is proved that $lr_{a}=0.001$, $lr_{c}=0.01$ performs best.
The choice of the discount factor significantly influences strategy selection and updates. According to the experimental results presented in Table, optimal performance is observed when $\gamma = 0.99$.
Furthermore, the escalation in the number of ATSF agents not only amplifies the training costs but also results in diminished resource utilization. Contrary to expectations, Completion Time exhibits an upward trajectory rather than a decline, as illustrated in Table. The optimal performance is achieved when the number of agents is set to 16.
The performance of GNN-weighted embedding is influenced by the maximum depth of root-leaf message passing, as illustrated in Table. The optimal performance is achieved when $max_{depth} = 8$

\begin{figure}
  \center 
  \includegraphics[scale=0.2]{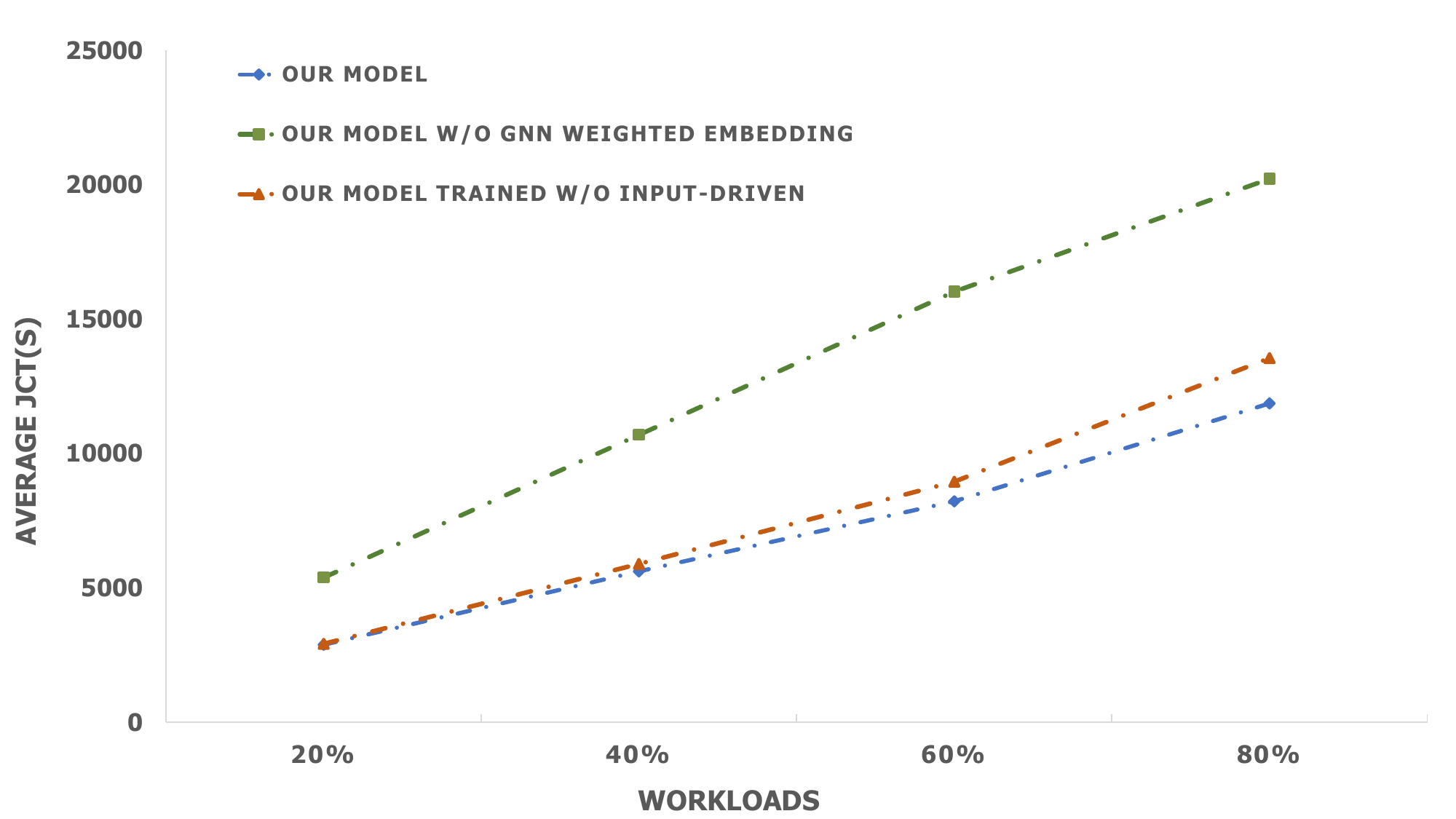} 
  \caption{Excluding any idea Raises The completion time of Our Model}
  \label{FIG:3}
\end{figure}

\begin{sidewaystable*}\Large
\centering
\caption{Comparing the essential parameters in our models}
\renewcommand{\arraystretch}{1.0}
\newcommand{\upcite}[1]{{\setcitestyle{square,super}\cite{#1}}}
\setlength{\tabcolsep}{25pt}
\resizebox{\linewidth}{!}{
\begin{tabular}{*{20}{lcccc}} % 控制表格的格式
\toprule
\multirow{4}{*}{Parameter} & \multicolumn{4}{c}{Workloads} \\
\cmidrule(lr){2-5}
& 20\% & 40\% & 60\% & 80\% \\
\cmidrule(lr){2-5}
& \multicolumn{4}{c}{Completion Time(s)} \\
 % \makecell[c]{ Turnaround times\\ Makespan \\ Completion time}\\
  \midrule
     \Large
    $Hd_{8}$      &  2308.88, 3765.29, 5454.33, 4603.68, 4091.86 & 5227.42, 7207.32, 8148.63, 7557.72, 5943.64 & 6795.88, 8831.78, 9249.31, 8793.75, 7600.94 & 9321.13, 9880.73, 13673.77, 12574.92, 11045.87 \\
     \Large
    $Hd_{16}$     &  2242.30, 3581.81, 3939.04, 3851.86, 3735.25 & 5447.74, 6306.80, 7715.63, 6888.21, 5833.12 & 6412.87, 7098.64, 7977.58, 8378.86, 8786.55 & 11090.00, 12021.70, 12754.72, 12342.31, 9415.35  \\
     \Large
    $Hd_{32}$     &  2188.87, 2802.00, 3186.84, 3377.98, 3768.57 & 5332.88, 4640.45, 5158.84, 5378.27, 5694.09 & 6749.84, 6841.51, 7709.10, 6329.86, 6207.16 & 9625.05, 11872.21, 9304.17, 9694.42, 9610.49 \\
     \Large
    $Hd_{64}$     &  5342.68, 5074.30, 6958.80, 7433.67, 9893.52 & 6816.07, 8167.97, 7891.74, 7391.98, 7268.89 & 7262.70, 8044.50, 11060.00, 9595.96, 9179.21 & 10699.50, 9375.16, 13858.51, 13051.80, 11720.34   \\
     \Large
    $\ell_{1}$    &  2179.27, 2075.69, 2568.34, 3074.51, 2196.00 & 5236.12, 4568.25, 6106.08, 6256.32, 4835.86 & 7891.56, 8956.33, 9056.14, 8562.31, 9152.38  & 11137.10, 8415.84, 12199.56, 12209.93, 10162.39    \\
     \Large
    $\ell_{2}$    &  1469.32, 1689.52, 2182.34, 2573.64, 2481.52 & 4107.60, 4028.61, 5001.47, 5458.52, 4684.79 & 5316.72, 6040.17, 6435.69, 6202.33, 6843.57  & 8497.71, 7335.64, 11929.01, 90428.72, 10629.80   \\
     \Large
    $\ell_{3}$    &  2277.02, 2589.45, 3514.21, 3189.28, 3078.19 & 4932.14, 4920.99, 6452.42, 6512.64, 4720.34  & 7895.36, 6825.13, 5428.56, 8945.61, 6321.85  & 9189.88, 7184.32, 12240.94, 11219.36, 11023.04   \\
     \Large
    $\ell_{4}$    &  1623.16, 2385.72, 4229.24, 2573.29, 2625.13 & 4417.15, 3883.37, 6120.50, 5762.25, 5115.53 & 5407.95, 5006.21, 7053.15, 6496.51, 6758.85   & 11325.78, 11302.71, 10902.96, 9512.82, 7767.59   \\    
     \Large
    $\gamma_{1}$  &  1646.11, 2158.94, 2231.45, 2014.85, 2561.49 & 3752.46, 4485.20, 3891.50, 3542.21, 4065.21  & 6017.95, 6897.34, 8966.95, 9079.94, 6113.27  & 10524.94, 8609.17, 12882.16, 12393.31, 11004.98  \\
     \Large
    $\gamma_{2}$  &  1717.56, 2347.89, 2356.51, 2871.68, 1896.71 & 4368.59, 4261.20, 5891.52, 6071.11, 4189.33  &  8047.32, 8897.44, 6996.11, 7064.00, 7113.98 & 10624.72, 9052.11, 15415.71, 12686.17, 12004.19   \\
     \Large
    $\gamma_{3}$  &  1622.41, 2081.78, 2275.46, 1536.41, 2349.11 & 3021.65, 2954.20, 3152.52, 3562.10, 3948.46 & 5845.61, 4851.20, 5421.00, 7054.79, 6044.52   & 8323.44, 8345.61, 6782.63, 12435.34, 10248.00   \\
     \Large
    $\gamma_{4}$  &  2746.53, 4695.34, 4410.97, 3096.77, 3006.00 & 4461.58, 4546.26, 6646.22, 6741.50, 4358.16 & 6200.55, 5047.10, 7662.99, 7127.06, 7361.41   & 8467.20, 11813.64, 6954.43, 11048.45, 10458.73   \\
     \Large
    $agent_{1}$   &  1621.59, 2258.79, 1600.13, 3520.14, 2564.30 & 5494.34, 6248.50, 8866.86, 7704.53, 5950.20 & 6577.89, 7563.73, 9063.18, 8115.68, 8211.48 & 8642.16, 7490.59, 12645.09, 11185.51, 10748.41 \\
     \Large
    $agent_{2}$   &  1676.37, 2884.56, 3521.78, 4042.71, 3314.00 & 6632.90, 7155.08, 8890.51, 8100.71, 6581.77 & 7536.93, 7030.96, 8502.53, 8414.43, 8431.23 & 9437.00, 8477.44, 13431.48, 11984.50, 11301.35   \\
     \Large
    $agent_{3}$   &  1593.99, 2074.62, 2243.58, 3010.30, 2247.12 & 4542.89, 4891.58, 6633.76, 6312.46, 4501.05 & 5404.44, 5294.85, 7630.42, 6112.51, 6958.38 & 8560.90, 7440.60, 12229.57, 11039.47, 10203.20   \\
     \Large
    $agent_{4}$   &  1610.39, 2997.38, 2975.55, 4455.60, 4569.50 & 4698.39, 5619.04, 6895.19, 7976.51, 5243.19 & 5845.64, 5772.16, 7536.13, 7573.31, 7377.92 & 8779.27, 7735.41, 11800.20, 1186.79,  10538.27  \\
     \Large
    $max_{depth1}$  &  1640.38, 3152.11, 2641.25, 3071.45, 3354.08 & 4556.03, 4857.93, 6544.60, 6061.85, 4625.18 & 5955.77, 5282.15, 7397.07, 7673.03,7577.13  & 8167.40, 8284.26, 11921.43, 11630.74, 10201.78   \\
     \Large
    $max_{depth2}$  &  1654.73, 2235.20, 3014.59, 3355.06, 2941.05 & 5846.91, 6566.16, 8338.26, 7939.66, 5993.95 & 6517.74, 6853.08, 9224.12, 8985.30, 8542.49 & 9033.55, 8401.43, 13814.02, 12215.98, 10884.45   \\
     \Large
    $max_{depth3}$  &  1560.99, 2036.01, 1865.35, 2546.03, 2786.13 & 4365.50, 4332.70, 7544.20, 6162.76, 4409.68 & 5809.55, 4780.40, 6343.50, 7267.32, 7821.14 & 8603.10, 6515.25, 11606.48, 10934.13, 10156.90   \\
     \Large
    $max_{depth}4$  &  1588.92, 1825.16, 3145.96, 3061.11, 2512.20 & 6611.40, 6397.10, 8477.39, 7679.18, 6587.36 & 7246.95, 7319.54, 7999.34, 8268.75, 8873.40 & 10015.60, 8340.66, 13108.93, 12149.37, 11196.74   \\
  \bottomrule    
  \end{tabular}
  }
  \label{tbl:table1}
\end{sidewaystable*}
Subsequently, we conducted additional ablation experiments to directly compare the impacts of parallelism control and input-driven methods on the retrained model using identical parameters. We excluded the parallelism control and input-driven approaches, leaving only our model, as depicted in Fig. \ref{FIG:3}. Through this comparison across different workload levels (20\%, 40\%, 60\%, and 80\%), we showcase the essential contributions of each key concept to the model's performance, as measured by completion time.\par 

\subsection{Algorithm Comparison}
The evaluation is conducted with baseline algorithms and deep reinforcement learning algorithms for comparison.

\begin{enumerate}[(1)]
\item RR \cite{elmougy2017novel}, The default scheduling method of Storm.
Round-robin scheduling is a basic process scheduling algorithm. 
Each process is assigned a fixed time slice, and when the time slice is exhausted, the system will switch to the next process in the ready queue.
\item DP \cite{sheikh2020dynamic}, Heuristic scheduling algorithm based on dynamic partitioning.
It dynamically allocates resources based on the characteristics of the task to improve overall performance.
This method can adapt to different task requirements and improve resource utilization.
\item PG \cite{kintsakis2019reinforcement}, A method for learning optimal behaviour through the interaction of an agent with its environment. Adaptable and able to cope with different environments and tasks.
\item GA \cite{xia2022multi}, Genetic algorithm is an optimization algorithm that simulates natural selection and genetic mechanisms.
In scheduling, it generates better scheduling solutions by simulating crossover and mutation of genes.
Strong global search ability and ability to find better solutions.
\item PPOC \cite{zhang2022ppo}, PPO-Clip is a policy-based reinforcement learning algorithm and a type of off-policy algorithm.
It improves stability by limiting policy changes per update.
\item Decima \cite{mao2019learning}, Baseline, Decima is the first method to combine deep reinforcement learning with cluster resource task scheduling.It provides a comparative benchmark that can be used to evaluate the performance of other algorithms.
\end{enumerate}

\begin{figure}
  \center 
  \includegraphics[scale=0.24]{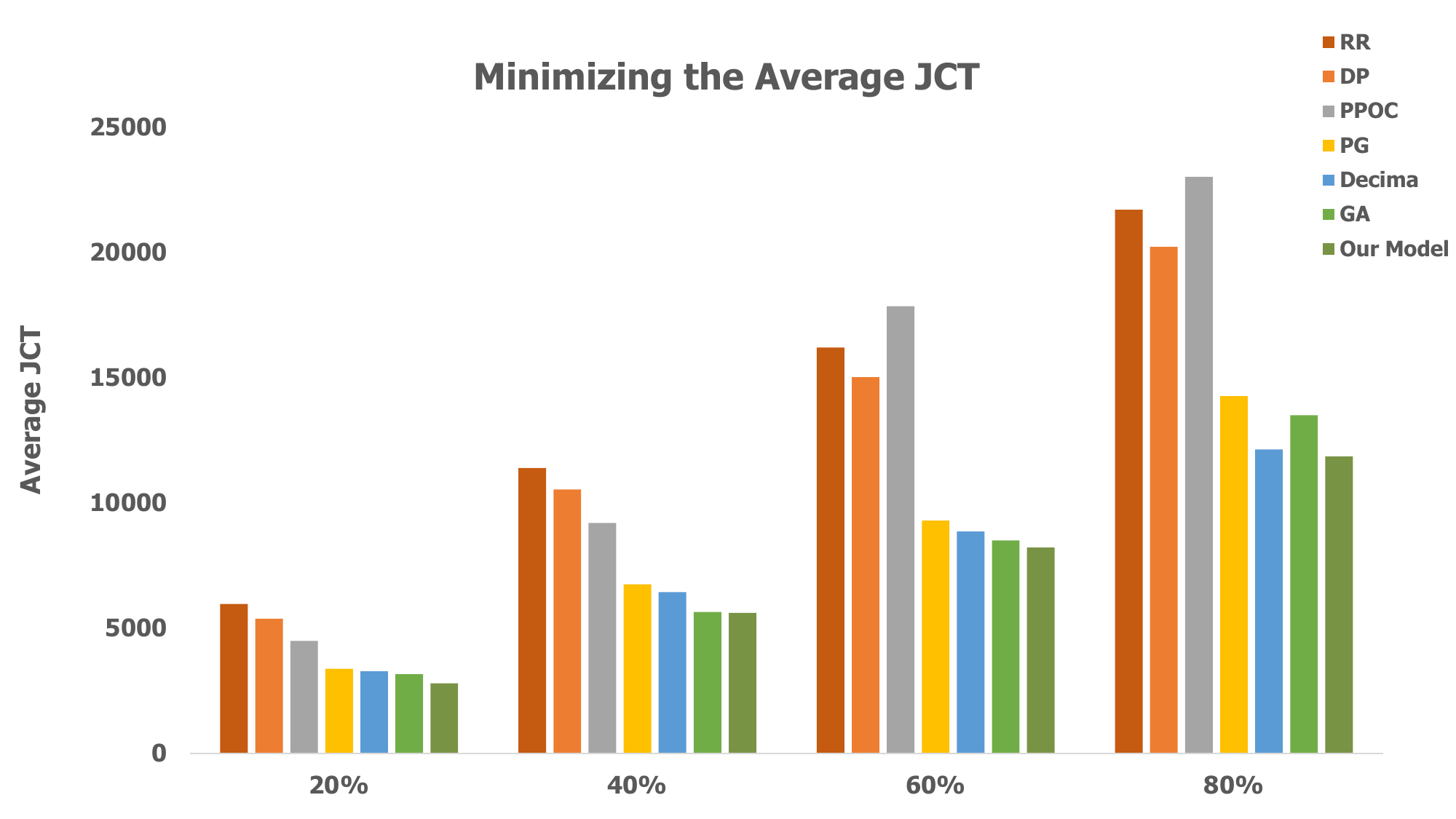} 
  \caption{Influence of Work Loads on completion time}
  \label{FIG:4}
\end{figure}

First, from the Storm cluster scenario where multiple streaming jobs coexist, we consider minimizing the completion time as the optimization goal to schedule these jobs. 
In the experiment, under the same conditions, we choose the number of Streaming jobs to be 1000, the arrival time of each job obeys the Poisson distribution, the average job arrival rate is $\gamma = 0.25$, and the policy network learning rate $\alpha$ is $1 \times 10^{-3}$, the value network learning rate $\beta$ is $1 \times  10^{-2}$, and then we compare our model with Decima\cite{mao2019learning}, PG\cite{kintsakis2019reinforcement}, RR\cite{elmougy2017novel}, DP\cite{sheikh2020dynamic}, GA\cite{xia2022multi} and PPOC\cite{zhang2022ppo}. 
In the TPC-H data set, the results of the completion time are shown in Fig\ref{FIG:4}, compared to Decima, our algorithm reduces the completion time by 17\%.
% \begin{figure}
%   \center 
%   \includegraphics[scale=0.24]{figs/test1.pdf} 
%   \caption{Comparison of completion time}
%   \label{FIG:4}
% \end{figure} 

In the Alibaba data set, as shown in Table\ref{tbl:table2}, To verify the effect of the model on different data sets. Our Model shows better performance in completion time. The main reason is that GNN-weighted embedding and improved Advantage Actor-critic can effectively balance peak and trough resource utilization and formulate efficient task executor matching strategies.

\begin{table}
\centering
\caption{Comparison of completion time with TPC-H and Alibaba data sets}
\renewcommand{\arraystretch}{1.0}
\newcommand{\upcite}[1]{{\setcitestyle{square,super}\cite{#1}}}
\setlength{\tabcolsep}{18pt}
\resizebox{\linewidth}{!}{
\begin{tabular}{*{10}{lcccc}} % 控制表格的格式
\toprule
\multirow{2}{*}{Model} & \multicolumn{1}{c}{TPC-H data sets} & \multicolumn{1}{c}{Alibaba data sets} \\
\cmidrule(lr){2-3}
& Completion Time(s)  & Completion Time(s)  \\
 % \makecell[c]{ Turnaround times\\ Makespan \\ Completion time}\\
  \midrule
  RR & 5983.432  & 6727.62 \\
  DP & 5388.605  & 5842.12\\
  PPOC & 4507.08  & 5942.3\\
  PG  & 3390.554  & 5908.2 \\
  Decima & 3293.671  & 3663.38 \\
  GA  & 3176.44 & 3032.85 \\
  Our Model & 2812.352  & 2911.59\\
  \bottomrule
  \end{tabular}
  }
  \label{tbl:table2}
\end{table}

Second, our research is directed towards examining the impact of the different workloads on completion time.
Fig. \ref{FIG:4} illustrates that as we incrementally raise the workload, even with the rise in the completion time, our model continues to exhibit strong performance under these conditions.

\begin{figure}
  \center 
  \includegraphics[scale=0.24]{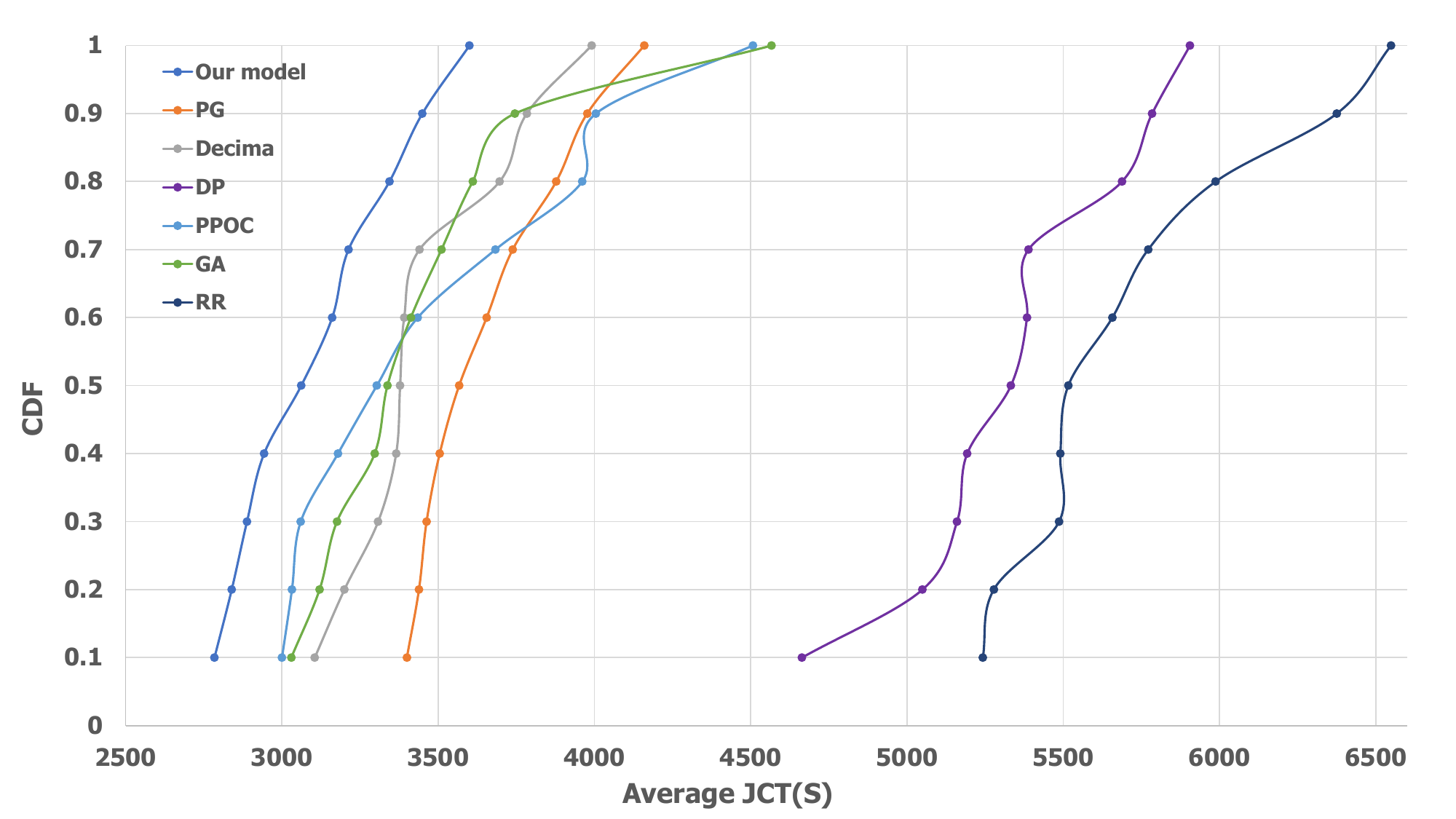} 
  \caption{CDF of completion time}
  \label{FIG:5}
\end{figure}

Next, our research is directed towards examining the impact of the job arrival rate, represented by the parameter $\gamma$ in the Poisson process, on the ATSF Agent. 
Fig. \ref{FIG:5} illustrates that as average reward accumulation in training iterations by ATSF agent.
Note that, the average rewards accumulating higher total reward implies the agent has learned a better policy which can optimize the actual objectives. 
Due to the randomness induced by the $\varepsilon$ -greedy and the instability of value networks sometimes the rewards drop for the ATSF agent.
% \begin{figure}
%   \center 
%   \includegraphics[scale=0.24]{figs/test5.pdf} 
%   \caption{Convergence of the ATSF algorithm}
%   \label{FIG:6}
% \end{figure} 

In the presented Fig. \ref{FIG:5}, we observe that due to the random amalgamation of stream jobs generated from TPC-H, these jobs exhibit both arbitrary interdependencies and variations in data volumes and execution durations. Consequently, investigating the cumulative distribution function (CDF) of the mean completion time becomes essential. Through experimentation, our model demonstrates superior performance and stability compared to alternative models. Specifically, when subjected to the same 20\% workload scenario, our model exhibits a completion time ranging from approximately 2750 to 3600. In contrast, the Decima\cite{mao2019learning} model showcases a completion time fluctuating between about 3100 to 4000, while the PG\cite{kintsakis2019reinforcement} model's completion time varies from around 3400 to 4200.\par 

Finally, We combined the TPC-H and Alibaba data sets in equal proportions to create a comprehensive streaming application collection, comprising 44 distinct applications. 
This integrated data set has been segmented into data partitions varying in size from 2 GB to 100 GB, catering to diverse application requirements.
Both the training and test sets for the experiments are randomly extracted from the data partition, and they share an identical internal distribution.
This data set will be utilized for parameter verification.

\section{CONCLUSION}
In this paper, in terms of the persistence, the periodicity and the spatial-temporal dependence of stream workload, a new Storm scheduler with Advantage Actor-Critic is proposed to improve resource utilization for minimizing the completion time. Our approach focuses on A new weighted embedding with a Graph Neural Network designed to represent features of a job comprehensively, which includes the relationships, the types and the positions of tasks in a job. 
An improved Advantage Actor-Critic integrating task chosen and executor assignment is proposed to schedule tasks to executors in order to better resource utilization. The outcome is a significant reduction in completion time for stream jobs, along with enhanced processing efficiency and fault tolerance in Storm clusters. Through experimental validation, we demonstrate that our model outperforms existing scheduling methods for Storm clusters, excelling in objectives like minimizing completion time and maintaining robustness.

\printcredits

\bibliographystyle{cas-model2-names}

\bibliography{cas-refs}


\bio{figs/pic1}
Author biography with author photo.
Author biography. Author biography. Author biography.
Author biography. Author biography. Author biography.
Author biography. Author biography. Author biography.
Author biography. Author biography. Author biography.
Author biography. Author biography. Author biography.
Author biography. Author biography. Author biography.
Author biography. Author biography. Author biography.
Author biography. Author biography. Author biography.
Author biography. Author biography. Author biography.
\endbio

\bio{figs/pic1}
Author biography with author photo.
Author biography. Author biography. Author biography.
Author biography. Author biography. Author biography.
Author biography. Author biography. Author biography.
Author biography. Author biography. Author biography.
\endbio

\bio{figs/pic1}
Author biography with author photo.
Author biography. Author biography. Author biography.
Author biography. Author biography. Author biography.
Author biography. Author biography. Author biography.
Author biography. Author biography. Author biography.
\endbio

\bio{figs/pic1}
Author biography with author photo.
Author biography. Author biography. Author biography.
Author biography. Author biography. Author biography.
Author biography. Author biography. Author biography.
Author biography. Author biography. Author biography.
\endbio
\end{sloppypar}
\end{document}